# Mobility enhancement and highly efficient gating of monolayer MoS$_2$ transistors with Polymer Electrolyte

Ming-Wei Lin[1], Lezhang Liu[1], Qing Lan[1], Xuebin Tan[2], Kulwinder Dhindsa[1], Peng Zeng[2], Vaman M. Naik[3], Mark Ming-Cheng Cheng[2], and Zhixian Zhou[1, a]

[1]Department of Physics and Astronomy, Wayne State University,
Detroit, MI 48201

[2]Department of Electrical and Computer Engineering, Wayne State University,
Detroit, MI 48202

[3]Department of Natural Sciences, University of Michigan-Dearborn
Dearborn, MI 48128

Abstract

We report electrical characterization of monolayer molybdenum disulfide (MoS$_2$) devices using a thin layer of polymer electrolyte consisting of poly(ethylene oxide) (PEO) and lithium perchlorate (LiClO$_4$) as both a contact-barrier reducer and channel mobility booster. We find that bare MoS$_2$ devices (without polymer electrolyte) fabricated on Si/SiO$_2$ have low channel mobility and large contact resistance, both of which severely limit the field-effect mobility of the devices. A thin layer of PEO/ LiClO$_4$ deposited on top of the devices not only substantially reduces the contact resistance but also boost the channel mobility, leading up to three-orders-of-magnitude enhancement of the field-effect mobility of the device. When the polymer electrolyte is used as a gate medium, the MoS$_2$ field-effect transistors exhibit excellent device characteristics such as a near ideal subthreshold swing and an on/off ratio of 10$^6$ as a result of the strong gate-channel coupling.

a) Author to whom correspondence should be addressed, electronic mail: zxzhou@wayne.edu

## 1. Introduction

Graphene has opened the tantalizing possibility of "post-silicon" high performance electronics because of its one atomic-layer thickness and extraordinarily high carrier mobility [1-4]. However, the lack of an appreciable bandgap in graphene poses a major problem for conventional digital applications. As a semiconducting analogue of graphene, single-layer $MoS_2$ has a direct bandgap of ~1.8 eV, which makes it a suitable channel material for low power digital electronics.[5] Similar to graphene, atomic layers of covalently bonded S-Mo-S unites can be extracted from bulk $MoS_2$ crystals by a mechanical cleavage technique due to relatively weak van der Waals interactions between the layers. However, the carrier mobility in monolayer and few-layer $MoS_2$ field-effect transistors (FETs) fabricated on $Si/SiO_2$ substrates are typically in the range of 0.1 -10 $cm^2V^{-1}s^{-1}$, which is not only orders of magnitude lower than that of graphene but also substantially lower than the phonon-scattering-limited mobility in bulk $MoS_2$ (which is on the order of 100 $cm^2V^{-1}s^{-1}$ ). [6-9]   Radisavljevic *et al*. have recently shown that the mobility of monolayer $MoS_2$ FETs can be improved to at least 200 $cm^2V^{-1}s^{-1}$ by depositing a thin layer of $HfO_2$ high-κ gate dielectric on top of $MoS_2$ devices, where the significant mobility enhancement was attributed to the suppression of Coulomb scattering due to the high-κ environment and modification of phonon dispersion.[10]  However, it is not clear to what extent the observed mobility increase can be attributed to the screening of charged impurities and phonon dispersion modification. On the one hand, a temperature-dependent electrical transport study of monolayer and few-layer $MoS_2$ FETs by Ghatak *et al*. suggests that the relatively low mobility in $MoS_2$ FET devices fabricated on $Si/SiO_2$ substrate is a channel effect, largely limited by the charge-impurity-induced electron localization.[7] On the other hand,   Lee *et al.* showed that the mobility in MoS$_2$ FETs fabricated on Si/SiO$_2$ substrate can be largely underestimated due to the Schottky barriers at the MoS$_2$/metal contacts.[11]

In this article, we report a simple method to fabricate high mobility (~10$^2$ cm$^2$V$^{-1}$s$^{-1}$) MoS$_2$ FETs by covering the devices with a thin layer of polymer electrolyte (PE) consisting of poly(ethylene oxide) (PEO) and lithium perchlorate (LiClO$_4$). The estimated room-temperature field-effect mobility of the monolayer MoS$_2$ FETs increases by up to three orders of magnitude upon adding the PE. To study the respective influence of the MoS$_2$/metal contacts and MoS$_2$ channel on the device characteristics, we fabricated multiple devices with different channel lengths on a single ribbon of monolayer MoS$_2$ with uniform width. Electrical characterization of these devices reveals that the PE-induced mobility enhancement can be attributed partially to the drastic reduction of contact resistance and partially to the increase of the channel mobility. The improvement of the channel mobility is likely due to the neutralization of the uncorrelated charged impurities on or near the MoS$_2$ channel by the counter ions in the PE.[12-14] Furthermore, we demonstrate for the first time that near ideal gate-channel coupling can be achieved in our PE gated MoS$_2$ FETs with the subthreshold swing approaching the theoretical limit of 60 mV dec$^{-1}$ at room temperature for metal–oxide–semiconductor field-effect transistors (MOSFETs).

## 2. Experimental Details

Monolayer MoS$_2$ flakes were produced by repeated splitting of MoS$_2$ crystals by a mechanical cleavage method, and subsequently transferred to degenerately doped silicon substrates covered with a 290 nm-thick thermal oxide layer.[6, 15] An optical microscope was used to identify monolayer (and few-layer) MoS$_2$ samples, which were further characterized by non-contact mode atomic force microscopy (AFM) and Raman spectroscopy. Figure 1a shows an atomic force microscopy (AFM) image of a typical monolayer $MoS_2$. From a line scan of the AFM image (figure 1b), we estimate that the $MoS_2$ sample is ~ 0.7 nm thick, corresponding to a single layer.[8, 10] Raman Spectra were collected using a Jobin–Yvon Horiba Triax 550 spectrometer, a liquid-nitrogen cooled charge-coupled device (CCD) detector, an Olympus model BX41 microscope with a 100 × objective, and a Modu-Laser (Stellar-Pro-L) Argon-ion laser operating at 514.5 nm. The laser spot size was ~ 1 μm in diameter and the laser power at the sample was maintained at low level (~ 200 μW) to avoid any heating effect. The Raman spectrum of the sample shows two peaks at 383.5 cm$^{-1}$ and 403 cm$^{-1}$ (figure 1c), which can be associated with the in-plane $E^1_{2g}$ and out-of-plane $A_{1g}$ vibrations of a monolayer $MoS_2$, respectively.[16]

FET devices of monolayer $MoS_2$ were fabricated using standard electron beam lithography and electron beam deposition of 5 nm of Ti and 50 nm of Au.[17] A PE was prepared in air by dissolving PEO and $LiClO_4$ in the 8:1 weight ratio in de-ionized water, and then drop casted onto the $MoS_2$ devices, where the PE gate electrodes were simultaneously patterned on the substrate along with the drain and source electrodes.[18] The PE-electrode was kept very close to the device channel; and the coverage of the PE was also limited to within an area of less than 100 μm around the channel and PE-electrode. Figure 2a shows a micrograph of a typical $MoS_2$ device with schematically illustrated PE. Electrical properties of the devices were measured by a Keithley 4200 semiconductor parameter analyzer in vacuum (~1 ×10$^{-6}$ Torr) and at room temperature (unless otherwise specified) both before and after adding the PE. The electrical measurements were conducted in both the Si back gate (with or without PE) and PE-gate configurations. As schematically shown in figure 4a, when a positive (negative) voltage is applied to a PE-gate-electrode near the device channel, negative(positive) and positive (negative) ions in the PE accumulate on the gate electrode and channel, respectively, forming electric double layers (EDL) at their interfaces with the electrolyte.[19]

## 3. Results and discussions

We first measured the electrical properties of several monolayer MoS$_2$ FET devices without PE and found a consistently low mobility between 0.1 and 1.5 cm$^2$V$^{-1}$S$^{-1}$, which is in agreement with the values reported in the literature.[6, 7, 20] Upon adding the PE, a significant mobility increase is observed in all devices. Figure 2b shows the low-bias linear conductivity defined as σ = L/W×I$_{ds}$/V$_{ds}$ versus back gate voltage in a typical monolayer MoS$_2$ device (device A) before and after adding the PE layer. Here L, W, I$_{ds}$, and V$_{ds}$ are the channel length (5.9 μm), channel width (0.6 μm), drain-source current, and drain-source voltage, respectively. The field-effect mobility estimated from the linear region of the transfer characteristics of the device using the formula μ= Δσ /($C_{bg}$Δ$V_g$) before and after adding the PE is ~0.1 cm$^2$V$^{-1}$s$^{-1}$ and ~ 150 cm$^2$V$^{-1}$S$^{-1}$, respectively. Here $C_{bg}$= 1.2 × 10$^{-8}$ F cm$^{-2}$ is the capacitance between the channel and the back gate per unit area ($C_{bg}$ = ε$_0$ ε$_r$/d; ε$_r$ = 3.9; d = 290 nm). Similar mobility improvement has been observed in monolayer MoS$_2$ FETs by Radisavljevic *et al.* upon depositing a thin layer of HfO$_2$ on top, which was attributed to the suppression of the Coulomb scattering due the high-κ dielectric environment and modification of phonon dispersion in MoS$_2$ monolayers.[10] However, the dielectric constant of the PE (ε = 5) used in this study is much lower than that of HfO$_2$.[21] Moreover, the mobility of the devices drops drastically upon cooling below the freezing temperature of the ions in the PE, ruling out dielectric screening as the dominant mechanism responsible for the mobility enhancement in our devices (see figure 3c and detailed discussion below).

A possible mechanism for the field-effect mobility improvement in our devices is the ionic screening effect. At any given back gate voltage, the free counter ions in the PE accumulate on the graphene surface to neutralize the uncorrelated charged impurities.[12-14] Two orders of magnitude increase of mobility has been previously observed in graphene FETs immersed in ionic solutions, which was attributed to the ionic screening of charged impurity scattering in graphene.[12, 14] Although the PE is expected to introduce additional charged impurities, studies on PE gated carbon nanotube and graphene FETs show that the mobility of these devices remains high (on the order of $10^3$ $cm^2V^{-1}s^{-1}$ ) upon adding PEO/LiClO$_4$ PE. [1, 8, 19, 22] One likely scenario is that the $Li^+$ and $ClO_4^-$ ions accumulated on the channel surfaces are correlated in contrast to the uncorrelated initial charged impurities near or on the channel surfaces. Even modest correlations in the position of charged impurities has been shown to substantially increase the mobility in graphene.[23] Therefore, the neutralization of the uncorrelated charged impurities on or near the $MoS_2$ surface by the counter ions from the PE is likely, at least partially, responsible for the orders of magnitude increase of the mobility upon addition of PE.

A second possibility is that the mobility of our $MoS_2$ devices without PE is substantially underestimated due to the presence of Schottky barriers at the $MoS_2$/metal contacts (the contact resistance was not excluded in calculating the mobility). Figure 3a shows the drain-source current ( $I_{ds}$ ) versus bias voltage ( $V_{ds}$ ) measured at different back gate voltages for the same $MoS_2$ device (device A) before depositing the PE. Although the device exhibits linear and symmetric $I_{ds}$-$V_{ds}$ dependence at low $V_{ds}$ (figure 3a inset), the $I_{ds}$-$V_{ds}$ behavior is non-linear and asymmetric at high bias-voltages. When the drain and source electrode connections are physically exchanged, the $I_{ds}$-$V_{ds}$ characteristics also change suggesting the presence of asymmetry and possibly non-negligible Schottky barriers at the contacts. It has been recently reported that the current flow in MoS$_2$ can be largely limited by the contact barriers leading to a significant underestimate of the mobility.[11] Modeling the $I_{ds}$-$V_{ds}$ characteristics of individual MoS$_2$ flakes with proper consideration of the contact barriers yields mobility values comparable to the estimated field-effect mobility in our PE-covered monolayer MoS$_2$ devices as well as that reported in HfO$_2$-covered MoS$_2$ devices.[10] Liu *et al.* have further demonstrated that the field-effect mobility of multilayer ( ~ 20 monolayers) MoS$_2$ FETs exceeds 500 cm$^2$/V.s due to the smaller bandgap (thus smaller Schottky barrier) compared to monolayer MoS$_2$.[24] Therefore, a substantial reduction of the contact barriers is also likely to significantly increase the slope of the transfer characteristics (dσ/dV$_g$ ), leading to a higher estimated field-effect mobility.

To shed more light on the origin of the PE-induced mobility enhancement in our MoS$_2$ FET devices, it is necessary to investigate the respective contributions of the MoS$_2$/metal contacts and the MoS$_2$ channel to the total resistance of the device at various gate voltages before and after adding the PE. In figure 3b, we plot the resistances of multiple FETs fabricated on the same monolayer MoS$_2$ ribbon with uniform width as a function of the channel length before adding the PE, where each resistance value is calculated from the slope of the $I_{ds}$-$V_{ds}$ characteristics in the low- bias linear regime as shown in the inset of figure 3a. It is obvious that the resistance increases nearly linearly with the channel length, from which the contact resistance is estimated to be 40 MΩ and 150 MΩ at $V_{bg}$ = 40 and 30 V, respectively. The scattering of data at $V_{bg}$ = 30 V may be due to the contact resistance variation among different devices. The channel resistance for the device with L = 5.9 µm ( device A ) is several times larger than its contact resistance at all gate voltages, suggesting that the field-effect behavior in our long channel devices is dominated by the channel instead of the contacts. This finding is consistent with that of Radisavljevic *et al.* [10] Upon applying the PE, the low-bias resistance of the device decreases to below 2 MΩ for $V_{bg} > 3$ V as shown in the figure 3b inset. In the linear region of the transfer characteristics (from which the field-effect mobility is estimated), the total resistance of the device (device A) with PE remains below 7 MΩ, which is significantly lower than either the contact resistance or the channel resistance alone without the PE. This finding shows that covering our single layer MoS$_2$ FETs with PE not only reduces the channel resistance but also lowers the contact barriers, both of which are critical to improving the field-effect mobility of the MoS$_2$ FETs. While the improvement of the channel mobility can be attributed to the neutralization of uncorrelated charged impurities, the reduction of the contact barriers could be due to the modification of the metal work function at the contacts by the PE. It has been shown that the adsorption of certain molecules on electrode-metal surfaces can induce a strong decrease in the work function.[25] As our MoS$_2$ devices are *n*-type, reducing the work function lowers the Schottky barriers at the contacts. [11] Due to the interplay of the variations in the channel mobility enhancement and contact barrier reduction, the resistance of the devices does not follow the linear dependence on the channel length, making it difficult to accurately extract the contact resistance in MoS$_2$ devices with PE.

Addition of a top dielectric medium has also been shown to increase the back gate capacitance by up to two orders of magnitude in graphene FETs, leading to an overestimation of the mobility when this dramatic capacitance increase was not accounted for.[26, 27] In order to rule out this possibility and further verify that the increase of $d\sigma/dV_g$ upon applying the PE was indeed due to the combined effects of contact resistance reduction and channel mobility increase, we show in Figure 3c the transfer characteristics of another monolayer MoS$_2$ device (device B) measured below and above the freezing temperature of the Li$^+$ and ClO$_4^-$ ions. The nearly two orders of magnitude lower $d\sigma/dV_g$ (which is proportional to the mobility) at 220 K than at 295 K is likely due to the freezing of both the Li$^+$ and ClO$_4^-$ ions inside the PEO polymer. Thus they are no longer able to dynamically neutralize the charged impurities on or near the MoS$_2$ channel as the charged impurities (including those in the SO$_2$ dielectric) move and redistribute during the back gate voltage sweeps. [23] [28] The dramatic decrease of the mobility below the freezing temperature of Li$^+$ and ClO$_4^-$ ions eliminates the possibility of overestimating the mobility in PE covered MoS$_2$ devices due to the dielectric-media-induced capacitance increase. To further rule out the possibility of PE induced capacitance increase as a major cause of the observed mobility increase, we also estimated the back gate capacitance with PE from the drain-source current versus PE-gate voltage ( $I_{ds}$-$V_{tg}$ ) measured at different $V_{bg}$ values as shown in figure 3d. When $V_{bg}$ is changed by 40 V, the $I_{ds}$- $V_{tg}$ curve shifts by 0.7-0.8 V along the $V_{tg}$ axis. Assuming that the PE-gate capacitance ( $C_{tg}$ ) is ~ 10$^{-6}$ F/cm$^2$ [19], the back gate capacitance with PE is estimated to be ~ 10$^{-8}$ F/cm$^2$ ( based on $C_{bg} = \Delta V_{tg}/\Delta V_{bg} \times C_{tg}$ ) consistent with the $C_{bg}$ value without PE, suggesting that the PE does not substantially influence the back gate capacitance.

In addition to serving as a contact-barrier reducer and channel-mobility booster, the PE can also be used as a gate material to substantially improve the gate efficiency by taking advantage of the large EDL capacitance at the PE/MoS$_2$ interface. In order to avoid chemically induced sample degradation, the applied PE-gate voltage was limited to a conservative range, in which the leak current was maintained below 200 pA. The Raman spectra of the single layer MoS$_2$ before adding the PE and after removing the PE (upon completion of all electrical measurements) are nearly identical, excluding the possibility of electrochemically induced sample degradation. Figure 4b shows the transfer characteristic of device B (the same device as in figure 3c) measured in the PE-gate configuration. The overall PE-gate dependence of the drain-source current closely resemble those reported in reference [10], where 30 nm of HfO$_2$ was used as the top-gate dielectric. The transfer characteristics remain essentially unchanged at different gate voltage sweeping rates. For a drain-source voltage of 300 mV, a current on-off ratio of $10^6$ is reached for $-2 < V_{tg} < 0.5$ V, and a subthreshold swing ( S ) of ~ 62 mV/decade is obtain. This S value is notably smaller than the S = 74 mV/decade reported in reference [10], and approaches the theoretical limit of 60 mV/decade, indicating that the gate efficiency of our PE-gated MoS$_2$ device is close to 1. Such a large gate efficiency can be attributed to the large EDL capacitance of the PE. The near ideal subthreshold swing along with the strongly linear dependence of $I_{ds}$ on $V_{ds}$ at various top-gate voltages (figure 4b inset) further suggests that the PE reduces the Schottky barriers to nearly ohmic.[29]

## 4. Conclusion

We have fabricated high mobility and high gate-efficiency monolayer MoS$_2$ FETs by simply adding PEO/LiClO$_4$ PE on top of the devices. A channel-length dependent study of the device characteristics suggests that the over $10^3$ time mobility increase upon adding the PE is due partially to the reduction of contact resistance and partially to the enhancement of channel mobility by the PE. We have also demonstrated excellent device performance with a nearly ideal subthreshhold swing (~ 60 mV/decade at room temperature) and an on/off ratio of $10^6$ in PE-gated devices.

**Acknowledgement**

This work was supported by NSF (No. ECCS-1128297). Part of this research was conducted at the Center for Nanophase Materials Sciences under project # CNMS2011-066.

**Figure 1.** (a) AFM image of a monolayer MoS$_2$ sample deposited on SiO$_2$ surface. (b) Line profile of the MoS$_2$ sample in (a). (c) A Raman spectrum of the same MoS$_2$ sample.

**Figure 2.** (a) An optical micrograph of a typical MoS$_2$ FET device with schematically sketched PE. (b) Conductivity of a representative MoS$_2$ FET (device A) measured in the Si-back gate configuration before and after adding the PEO/LiClO$_4$ PE.

**Figure 3.** (a) Current-voltage characteristics of device A measured at various gate voltages before adding the PEO/LiClO$_4$ PE. (b) Resistance of FET devices fabricated on the same Monolayer MoS$_2$ (where device A was fabricated) as an function of channel length measured at different back gate voltages. (c) Conductivity as a function of back gate voltage of device B measured at temperatures below and above the freezing temperature of the ions in the PE. (d) Drain-source current versus PE-gate voltage of a MoS$_2$ FET device (device C) measured at $V_{ds}$ = 100 mV and various back gate voltages. The inset in (a) is the low-bias linear region of (a); and the inset in (b) is the Resistance of device A as a function of back gate voltage after adding the PEO/LiClO$_4$ PE.

**Figure 4.** (a) A schematic illustration of the working principle of PE-gated MoS$_2$ FETs. (b) Drain-source current versus PE-gate voltage of a Monolayer MoS$_2$ FET device (device B) measured at different drain-source voltages. The inset show the current-voltage characteristics at different PE gate voltages.

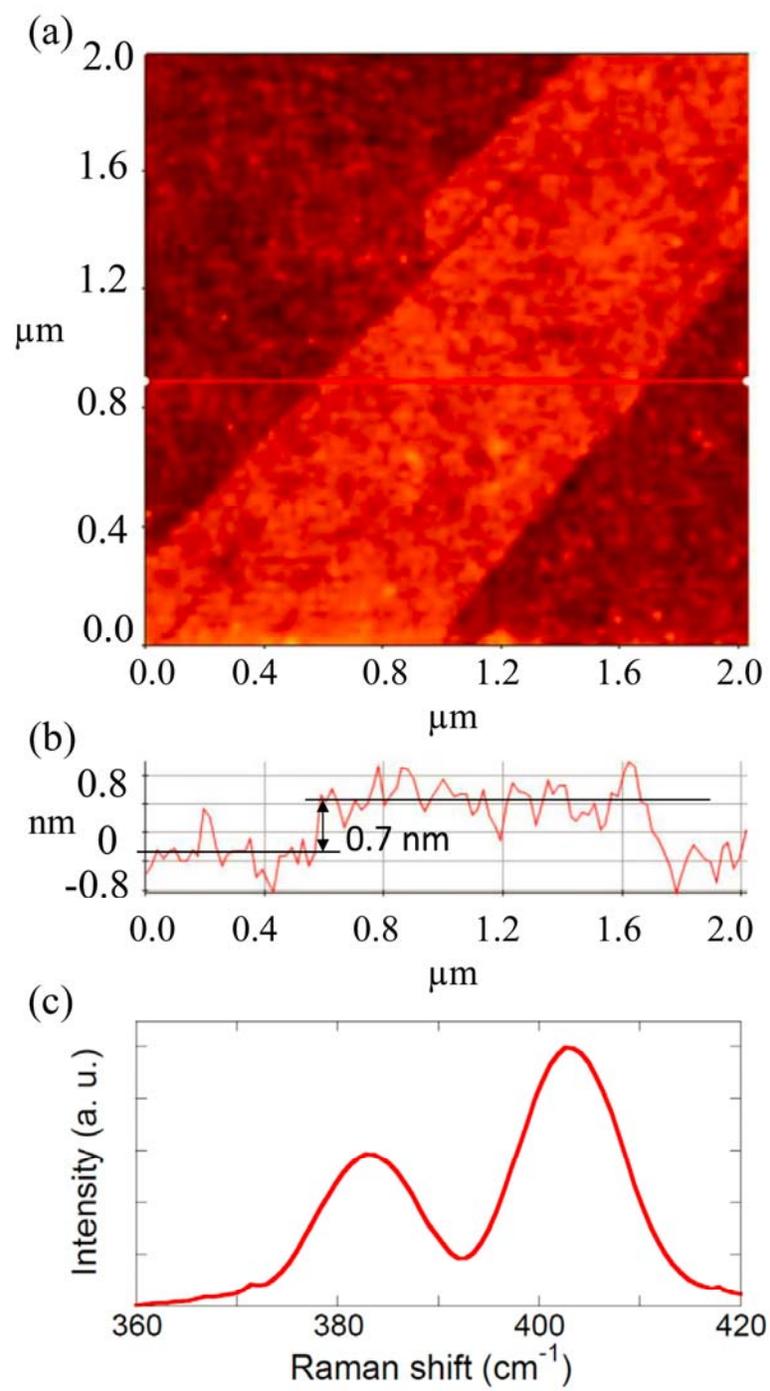

Figure 1

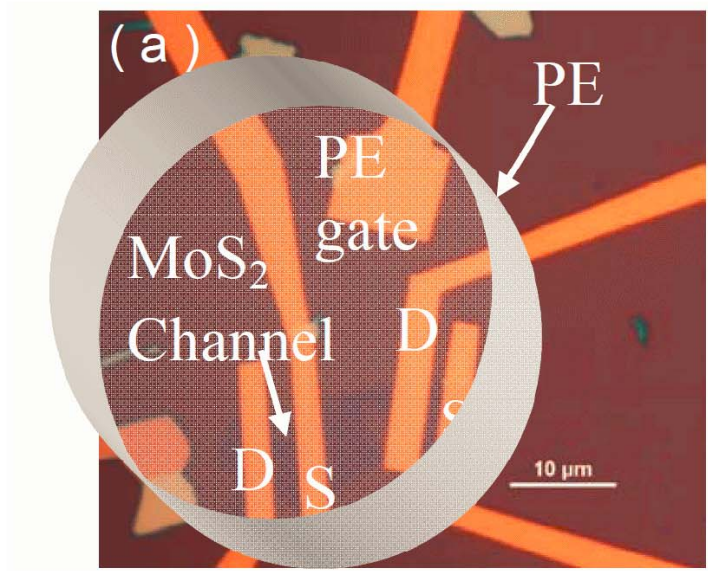

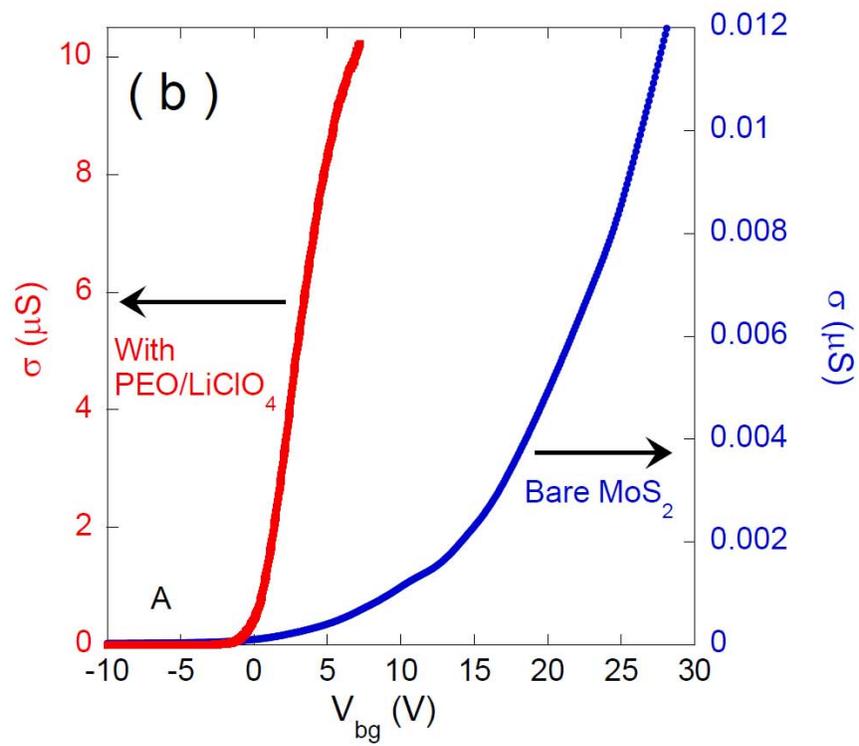

Figure 2

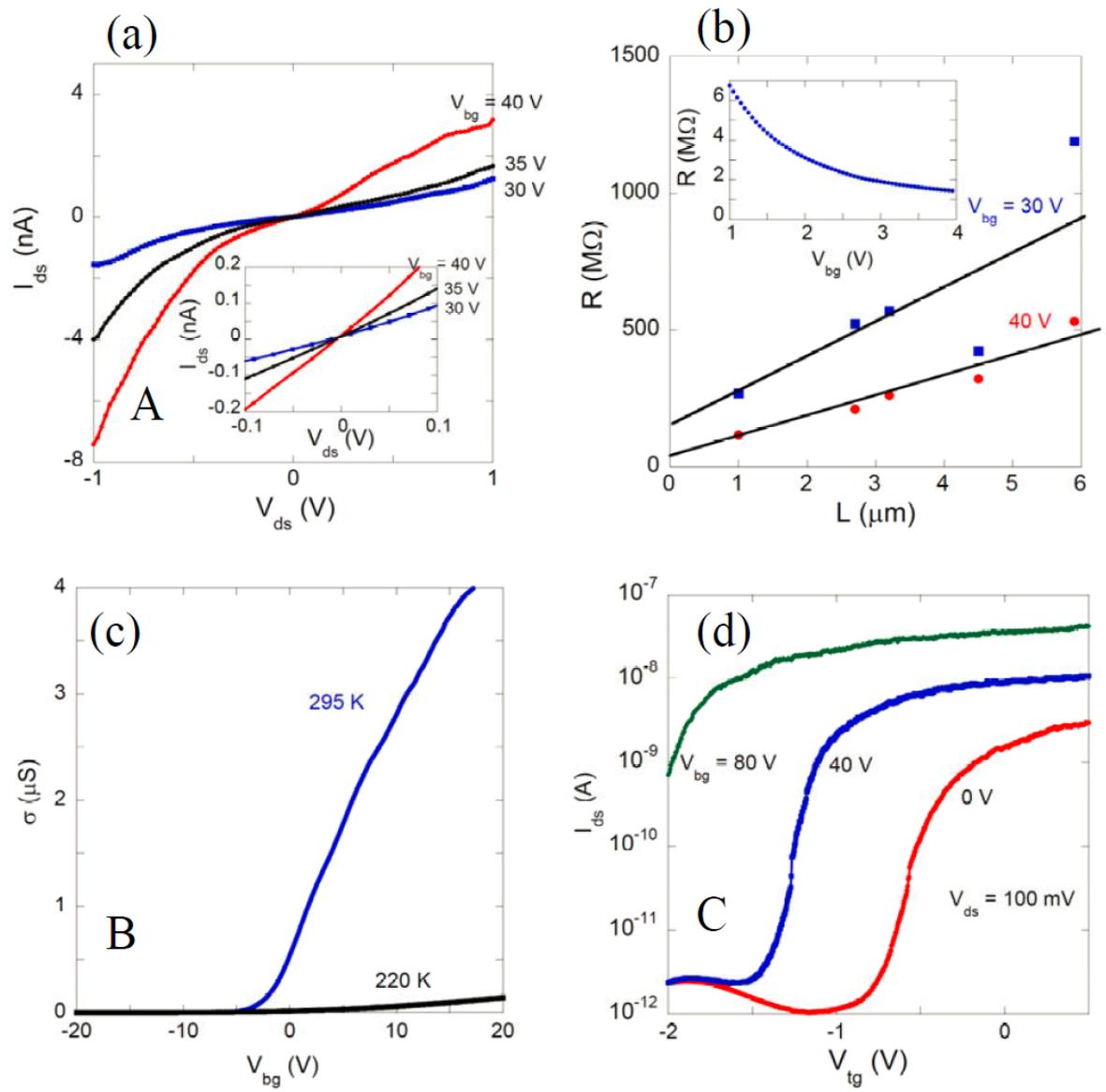

Figure 3

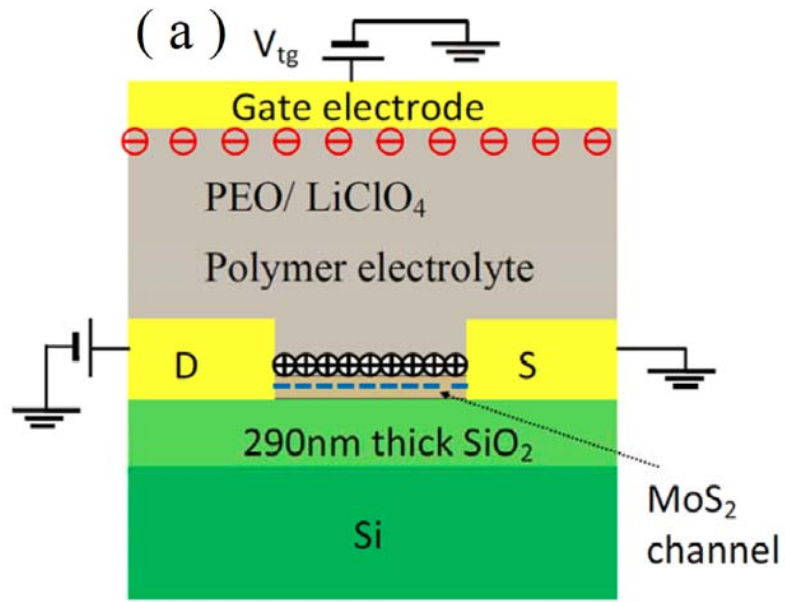

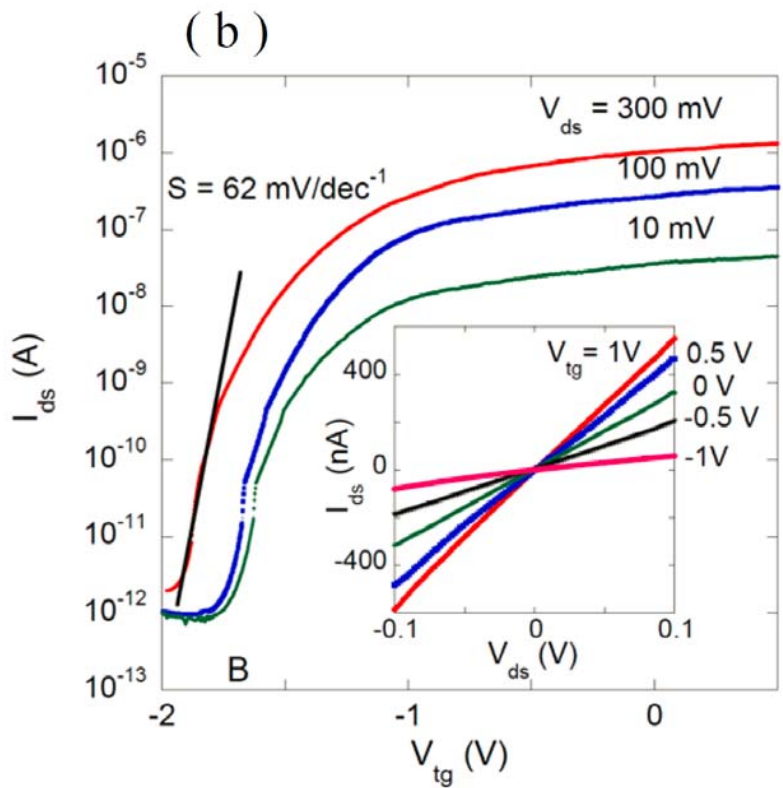

Figure 4